# 基于区块链的联邦学习数据确权机制研究

郭韧[1]，程小刚[2]
1.华侨大学 工商管理学院，福建 泉州 362021
2.华侨大学 计算机科学与技术学院，福建 厦门 361021

**摘要：**联邦学习可以解决分布式数据挖掘与机器学习中的隐私保护问题，而如何保护参与联邦学习的各方对各自数据的所有权、使用权和收益权是个重要的问题；提出一种基于区块链和智能合约的联邦学习数据确权机制，利用去中心化的区块链技术将各参与方的贡献大小保存于区块链之上，并通过区块链进行联邦学习成果的利益分配；并在区块链的本地模拟环境中模拟设计实现了相关的智能合约和数据结构，初步论证了方案的可行性。

# Research on Data Right Confirmation Mechanism of Federated Learning based on Blockchain

Xiaogang Cheng[1], Ren Guo[2]
1. College of Computer Science and Technology, Huaqiao University, Xiamen 361021, China
2. College of Business Administration, Huaqiao University, Quanzhou 362021, China
**Abstract:** Federated learning can solve the privacy protection problem in distributed data mining and machine learning, and how to protect the ownership, use and income rights of all parties involved in federated learning is an important issue. This paper proposes a federated learning data ownership confirmation mechanism based on blockchain and smart contract, which uses decentralized blockchain technology to save the contribution of each participant on the blockchain, and distributes the benefits of federated learning results through the blockchain. In the local simulation environment of the blockchain, the relevant smart contracts and data structures are simulated and implemented, and the feasibility of the scheme is preliminarily demonstrated.

## 1.引言

为解决信息孤岛问题，联邦学习被提出[1]，即在保护隐私的基础上进行数据挖掘和机器学习，各参与方在本地用自己的数据对神经网络模型进行训练，然后将训练后所得的神经网络梯度传送给汇总方，汇总方汇聚各方的局部梯度数据，得到总的模型在分发给各参与方。也即各参与方向外传送的信息是梯度而不是原始数据，从而一定程度上保护了各参与方的隐私。但此种基于分散数据的机器学习方式也会受到一定的挑战和攻击，如梯度反演攻击[2]，即从共享的梯度中反向推出各参与方的隐私数据，对于此种攻击可利用经典的DP（Differential Privacy，差分隐私）[3,4]进行应对，即在数据中加入扰动和噪声来保护隐私，显然加入的噪声越多就可以越好的保护隐私，但训练出的模型效用准确性就大大降低，所以要考虑隐私和效用的平衡；另一解决方案是同态加密[5]，缺点是同态加密开销过大，使得机器学习计算和时间成本增加；此外还有拜占庭攻击[6]，或称投毒攻击[7]，即部

基金项目：福建省社会科学规划项目(FJ2024B088)；福建省高校以马克思主义为指导的哲学社会科学学科基础理论研究项目 (FJ2024MGCA028)；教育部人文社会科学研究专项项目（24JD710008）

分参与方或节点故意，或非故意由于本身的软硬件故障，提供错误的训练参数或数据，从而影响最终的学习模型的效果。应对此种攻击需要汇聚方能够精准的识别出恶意行为并将其排除在外，从而降低其对学习模型的破坏。联邦学习自被提出以来，被广泛应用于各个领域如恶意流量监测[8]、医学[9]、船舶油耗[10]、食品安全[11]、工业制造[12]、图像的语义分割[13]等等。

在今天大数据时代，如何保护数据所有者的权益也是一个重要研究课题；数据拥有者对数据的权力包括知情权、决定权、查阅权、复制权、更正权、删除权等[14]；[15]中分析了我国出台的"数据二十条"的内涵，提出数据确权后续建设要点和方向；[16]以数据杀熟为例探讨了如何避免数据滥用，保护用户利益；[17]探讨了构建数据交易平台的关键技术如区块链、智能合约等；[18]中提出数据确权制度可刺激数据供给，提高数据质量；[19]中探讨了人工智能生成内容的确权模式和保护方式。

区块链技术中，各参与方在没有中心服务器的条件下，共同维护一个全局唯一的即记账本，从而记录在此记账本上的内容就具有权威性和不可抵赖性，因其得到了全体成员的认可和背书[20, 21]。在新一代信息技术环境下，"ABCD5"（即人工智能、区块链、云计算、大数据和 5G 技术）已成为情报工作的技术支撑体系[22]。[23]中基于区块链构建数据治理平台，明确政府数据协同治理过程中的数据归属权等问题。在[24]中，利用区块链技术来防止中心服务器的恶意行为和其单点失效问题，即各参与方在把中间的梯度值传递给中心聚合服务器的同时，也把信息存储在区块链上，这样即使中心服务器出现问题，也可以比较简单的基于区块链中的数据恢复学习过程，而不是全部从头再来；为避免传统区块链共识机制所浪费的算力和带宽，[25]中设计了基于联邦学习的共识协议，实现了联邦学习与区块链技术的优势互补；[26]中结合区块链和联邦学习技术，提出了一种食品全程风险信息可信共享模式，将联邦学习融入区块链架构中构建可信共享模型。

本文研究如何在联邦学习中进行数据确权，保护联邦学习中各参与方的权益，提出一种基于区块链和智能合约的数据确权机制；参与方在参与分布式联邦学习过程中，同时将自己的贡献值通过智能合约上传到区块链上去，此贡献值信息必须要得到 Dealer（联邦学习的发起者与聚合方）的确认才会生效；Dealer 在得到各参与的本地训练结果之后，检查合格过后要通过智能合约对贡献值进行确认，然后合并各参与方的本地训练结果，得到最终模型产品并进行发布；产生的收益可通过智能合约根据各方的贡献值大小进行分配；此方案可充分保护各方的利益，吸引掌握数据的各方参与项目的开发，因本方案在保证其自身隐私的前提下，还能获得最终产品收益的分享；使用本方案可更好更充分利用分布在社会各单位、个人手中的分散数据，本地、个人的数据量少价值不大，但通过参与大项目进行合作分享可发挥这些分散数据的最大价值，合作解决一些大的社会、科学问题研究等，而且由于可参与最终产品的收益分配与隐私保护的支持，解决以往分散数据利用的痛点，可刺激各单位个人积极参与，形成良性循环。

比如以流行音乐创作工具为例，大量的歌曲版权属于各个唱片公司，如果某个人工智能公司想要开发一款流行音乐歌曲、歌词创作工具，那么显然能获得越多歌曲数据进行训练的话，就能使产品质量更好，越能在激烈的市场竞争中脱颖而出；而对各参与的唱片公司来说要有相关机制保证其参与之后能获得相应的报酬；利用我们的方案可较好的解决此种情形的问题，开发公司在区块链上发布相关的智能合约，而有意参与的唱片公司可联系开发公司获取相关技术细节，根据自己的数据进行联邦学习中的本地训练，而后把局部结果传给开发公司，同时在区块链上写入自己的贡献值大小作为日后产品收益的分配依据；开发公司收到各方的局部训练结果后，进行检查并在区块链上进行确认，并聚合生成最终软件产品和发布；吸引用户并取得收益之后，也是通过智能合约根据各方的贡献值大小进行分配。另一个例子是本方案可适用于各种疾病的研究，比如糖尿病、癌症、阿尔兹海默

症等等；大量的病人病历等数据当然保存于各个医院之中，出于隐私保护的相关法律法规规定，医院当然不能公开这些数据用于科学研究，利用差分隐私对数据加入适量噪声后公开数据是一种解决方案，但显然隐私保护和数据的准确性和有效性是矛盾的，需要做到平衡；而我们的方案提供了另一种解决方案，保有数据的各个分散的医院参与一个全社会联合的疾病研究联邦学习项目，各医院在本地训练局部模型，然后传给 Dealer 局部训练的神经网络模型，再由 Dealer 合并生成最终的疾病辅助诊断治疗系统，这样可以更好保护个人隐私，又可以充分利用分散在社会各个单位之中的分散数据，进行科学攻关研究，并且参与联合研究的单位由于获得了后续产品收益的分享权，这对激励社会上用户源数据的主体参与这种联邦学习项目。

本文安排如下，第 2 节中给出一些初步知识；具体的联邦学习的数据确权方案的设计和思路在第 3 节；第 4 节中基于本地区块链仿真环境具体实现了我们设计的方案，并给出相关智能合约的核心代码，初步验证了本文方案的可行性；第 5 节结束语，并给出未来值得进一步研究的课题。

## 2.初步知识

联邦学习要解决的问题是信息孤岛问题，在数据挖掘和机器学习中，把分散的各处的数据集中到一起在进行挖掘和学习当然可以达到最好的学习效果，但这会涉及到隐私保护的问题，容易造成个人的隐私数据泄露；为此 Google 公司在 2017 年提出了联邦学习的概念和框架，其核心思想在于各局部数据拥有者利用自己的数据，在本地训练一个局部的神经网络模型，然后把训练好的局部神经网络模型的参数传输给集成方，这样在互联网上传输的是神经网络模型参数，而不是隐私数据，从而在一定程度上保护的个人隐私。根据各方持有的数据集之间的关系不同，联邦学习又可分为横向联邦学习、纵向联邦学习和联邦迁移学习；横向联邦学习指的是各方的数据特征相同但样本 ID 不同，比如不同的医院对疾病治疗的合作研究，各个医院病人不同，但数据特征如各种病人的特征参数如血压、血糖等相同或类似；纵向联邦学习指的是各方数据特征不同，但样本 ID 信息相同，比如比如政府部门对失信人员的监测和预警等，会涉及各个部门如银行、保险、法院等等，某人的 ID 一样，但显然在各个部门中所保存的信息是完全不一样的；而迁移学习中，各方的数据特征和 ID 信息都不同，比如中国的电商平台和外国银行之间进行合作研究，那么双方的数据 ID 和特征都不同。

区块链的基本核心思想在于去中心化，解决的问题是分散的各方如何在一个没有中心可信的权威机构的情形下，如何达成一致；如典型的应用 Bitcoin 就是各方共同维护一个唯一的账本（Ledger），记录所有的比特币交易数据，从而防止比特币的 Double Spending 等问题；常用的实现意见一致的方法是 PoW（Proof of Work），即各方同意将长度最长的区块链作为唯一正确的账本，而延长区块链的方式是解决一个相对比较困难的问题（如 Hash 函数输出的前若干位为 0 等），这样只要整个系统中诚实用户的总的计算能力大于攻击者，就可保证系统的正常运行；PoW 显然的缺点是在用算力证明的过程（即“挖矿”）中要耗费大量的 CPU 或 GPU 计算能力和电力等资源，所以也出现了其他的方式如 PoS（Proof of Stake），权益证明，即拥有的权益越多，则获得记账权的机会就越大；DPoS （Delegated POS），委托权益证明， 是由 PoS 演化而来的。持币用户通过抵押代币获得选票，以投票的方式选出若干的节点作为区块生产者，代表持币用户履行产生区块的义务。DPoS 与公司董事会制度相似，让持币用户将生产区块的工作委托给更有能力胜任的专业人士去完成，同时也能享受参与出块获得的奖励。

智能合约是一段保存于区块链上的代码，一旦某个条件满足，就会触发合约中的条款，自动执行相关代码，不需要人为操控；比如体育博彩的例子，某人向博彩公司购买了

某场足球比赛竞猜结果的彩票，结果揭晓后有可能出现博彩公司毁约失约、耍赖的情况，使得彩票购买人承受损失，而如果把这些约定编成代码，录入区块链中成为智能合约，那么一旦触发约定时的条件，就会有程序自动来执行，从而防止上述的违约行为。

**3.设计与实现**

我们的基于区块链与智能合约的联邦学习数据确权与收益分配机制运行如下（参见图 1）：

1. Dealer 发出要进行离散联邦机器学习的帖子，比如糖尿病的发现与治疗、音乐创作 AI 工具的开发等等；

2. 拥有相关资源和数据的参与方，可根据 Dealer 的帖子内容（比如 Dealer 的信誉实力、产品成功的几率、收益分成等等），并结合自身的实际情况决定是否参与联邦机器学习的过程；

3. 如果某方决定参与，则参与方向区块链上写入一条消息表明自己的参与意向和贡献大小，比如贡献的数据量（病人的病历数、音乐乐谱和歌曲的数量等）；然后根据联邦学习的机制在本地根据 Dealer 发布的神经网络模型用这些数据进行训练，并把训练的结果，及神经网络的梯度数值等传递给 Dealer；

4.上述参与方在区块链中写入的消息不能直接作为最终贡献大小的确认，而要经过 Dealer 的检查和确认，即 Dealer 在收到参与方发来的部分训练结果梯度值之后，要进行检查，无误后确认参与方写入的消息，来确认此方的参与和贡献；

5. Dealer 收集到各参与方的局部训练结果，检查并在区块链中确认之后，将各方的局部结果汇总形成最终的系统产品，并向外发布；同时统计汇总各参与方的贡献额度和比例，作为日后产品收益分配的依据；

6. 训练后的产品，如疾病诊断工具、音乐 AI 创作工具等，由 Dealer 向外发布，供用户有偿使用，按次或包月包年使用等等，得到的收益则按各方同意的机制分配；

7. 产品运行一段时间后，如果某个参与方对收益或分成等方面不满意，也可以选择退出此产品或系统，这可以通过参与方与 Dealer 运行相关“忘记”（Unlearning）[27]的算法来将此参与方的数据和贡献从产品中删除，并在区块链中进行相应的记录；表明本产品没有此参与方的贡献，并且此后此参与方也不参与产品收益的分成等等;

8. 收益的分配可利用智能合约自动进行利益分配，比如可以设定一个阈值（如一百万元），如果达到则自动触发收益分红的操作；或者收益分配由外部事件触发（比如股东大会决议或国家法律规定等），则可以利用区块链的预言机（Oracle）机制来连接区块链和外部世界从而触发收益分配。

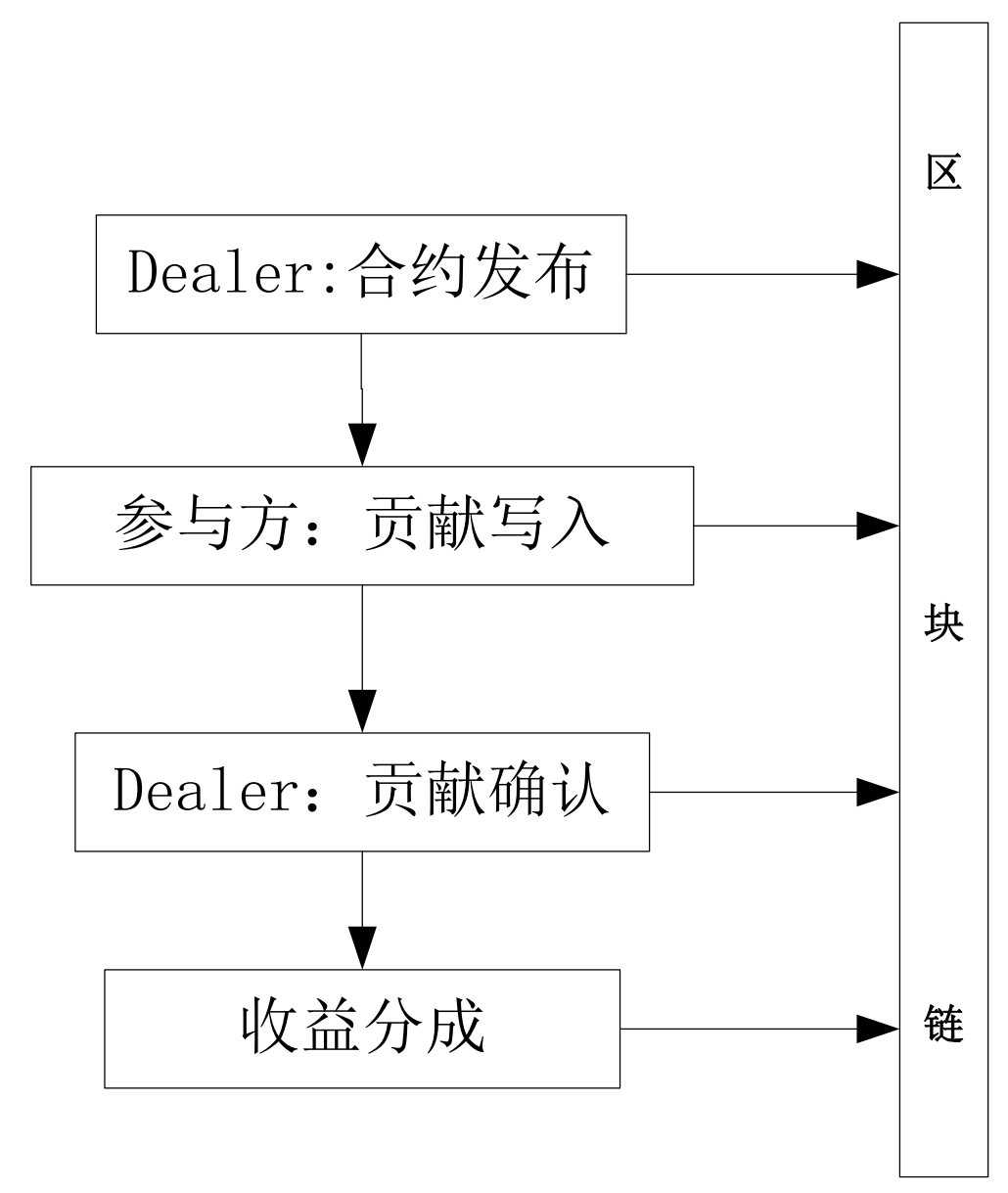

图 1、基于区块链的联邦学习数据确权机制示意图

## 4. 实现与仿真

### 4.1 实验环境搭建

仿真实验采用 Ganache 区块链本地仿真环境，MetaMask 小狐狸钱包和 Remix 智能合约开发、编译和部署 IDE；采用 Solidity 语言开发智能合约，网页前端采用 ether.js 库同智能合约进行交互。实验环境搭建和实验步骤如下：

1. 安装好 Ganache 区块链本地仿真环境之后，默认会创建十个账户，每个账户会有 100 个 ETH 以太坊币的余额：

| ADDRESS | BALANCE | TX COUNT | INDEX |
| --- | --- | --- | --- |
| 0×01d523e363837328a7dEb853DA3EAF70a5FAB35B | 100.00 ETH | 0 | 3 |
| 0×6d914E4A2D1de39564b33ec22704Cb5839c6Db02 | 100.00 ETH | 0 | 4 |
| 0×5390e0C9fb15cDF46A4fd244CdE33202131D81e2 | 100.00 ETH | 0 | 5 |
| 0×416b34954dc2d2DE87e33616B502Fb7E87934311 | 100.00 ETH | 0 | 6 |
| 0×0F04A498918C72fA89114fD01FEa403B84D0bDf6 | 100.00 ETH | 0 | 7 |
| 0×AB23bc3bD69BEd28945E40D9bdeD010708467C00 | 100.00 ETH | 0 | 8 |
| 0×2CA3CDf4b7cf08f9b6E3167807D2406904e946A3 | 100.00 ETH | 0 | 9 |

图 2、Ganache 本地账户

2. 安装 MetaMask 小狐狸钱包，并把 MetaMask 钱包同某个上述的 Ganache 账户连接起来，这样该账户后续可利用小狐狸钱包进行交易，如转账、交易，以及部署合约、合约调用等需要支付的 Gas 燃料费（即手续费）等；

3. 搭建好以上模拟环境和钱包之后，即可进行智能合约的开发和部署；如可利用 Remix 开发环境和 Solidity 语言进行开发：

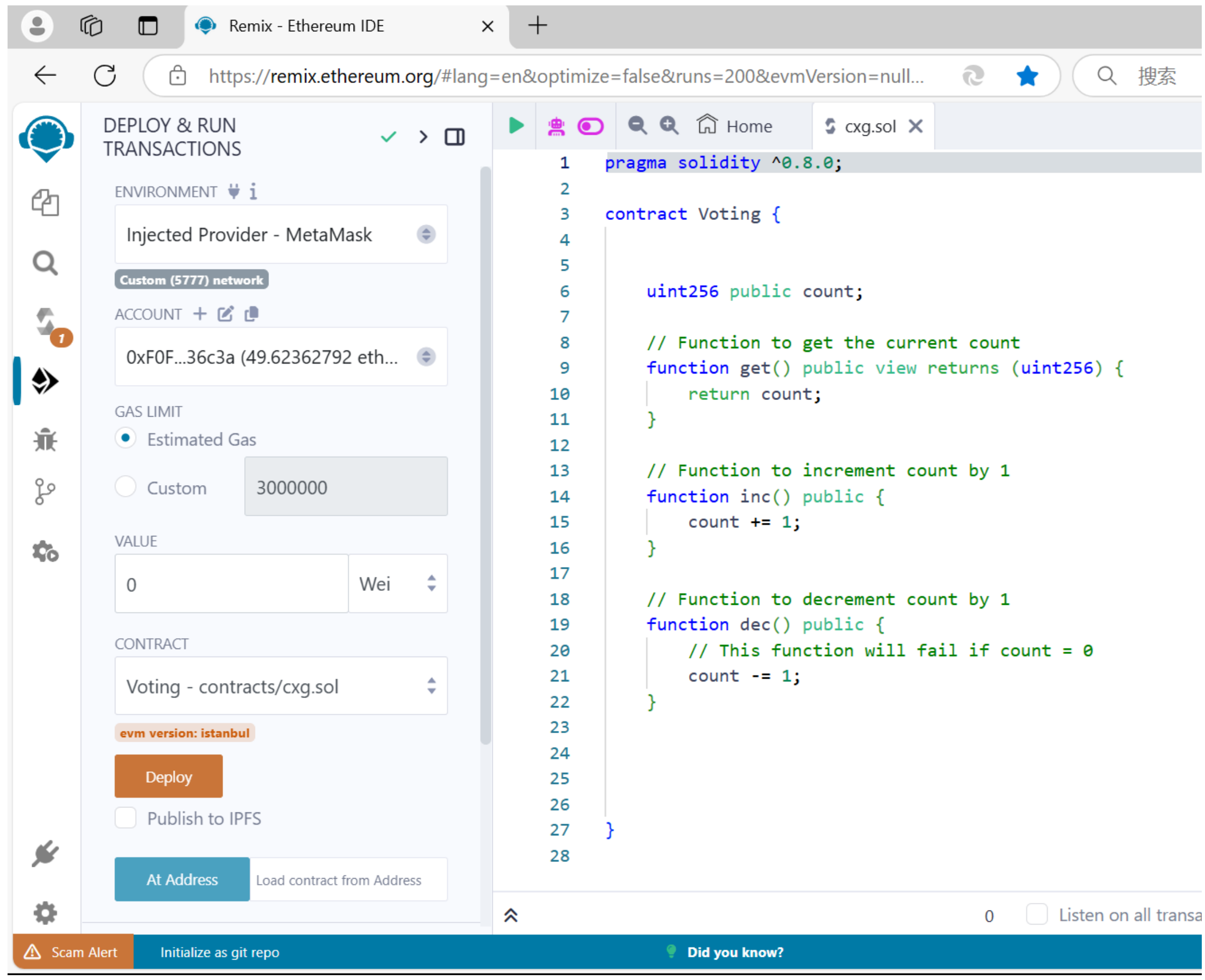

图 3、Remix 智能合约开发和部署平台

合约开发好之后，可以进行编译并通过 MetaMask 部署到本地的 Ganache 区块链上，而且部署合约是要支付一定的燃料费，会从上述绑定的 Ganache 账户中的余额中扣除：

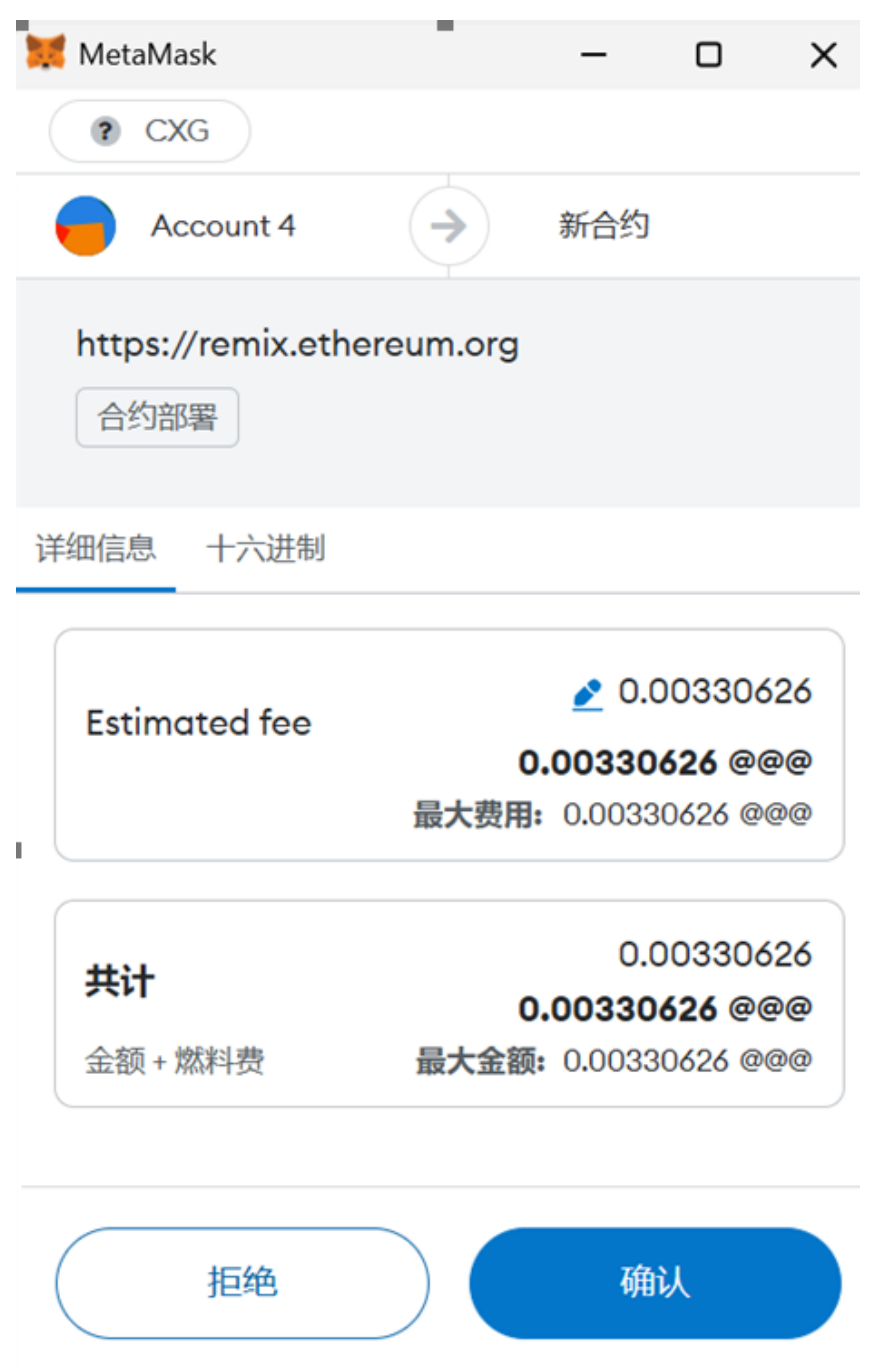

图 4、MetaMask 合约部署确认

4. 合约部署完成后，即可通过 Web 方式利用 ether.js 库对合约进行调用，首先连接小狐狸钱包并创建提供者 Provider （是一个连接以太坊网络的抽象），以及 Signer（是以太坊账户的抽象，可以用来签名消息和交易），然后可以连接合约 Contract：

```
let accounts = await window.ethereum.request({method:'eth_requestAccounts'});
document.getElementById("account").innerText=accounts[0];
let provider = new ethers.BrowserProvider(window.ethereum);
const signer = await provider.getSigner();
let contract = new ethers.Contract(cAdd,cABI,signer);
```

其中 cAdd 是合约地址如：

```
const cAdd="0x032C49108D988059dF3cc452Ad8cfDf997cadaA3";
```

cABI 是调用合约的应用程序二进制接口，提供与不同合约组件交互的规范；然后即可调用合约中的函数如：

```
let result = await contract.get()
```

对于此种查询类的操作，不会改变区块链的状态就不需要支付 Gas 燃料费，否则如果函数调用会改变区块链的状态，就要支付燃料费，也就需要小狐狸钱包应用中去批准和支付：

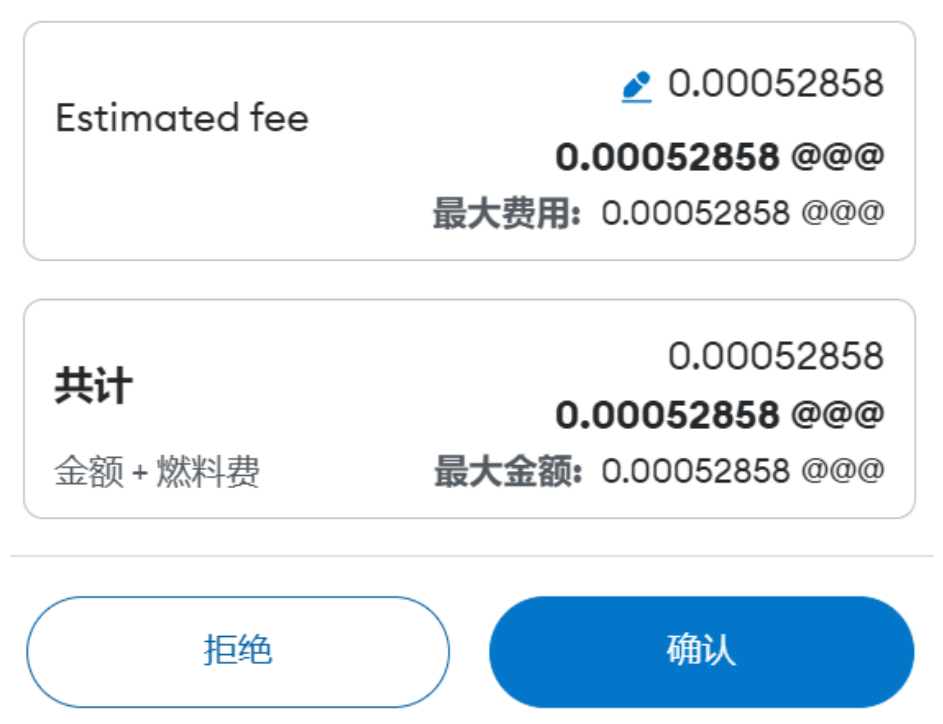

图 5、MetaMask 函数调用燃料费确认

如果用户选择拒绝，则此调用不会执行，确认后才会执行调用，并从用户账户中扣除相应的燃料费。合约部署和需要支付燃料费的区块链调用会在 Ganache 留下交易记录：

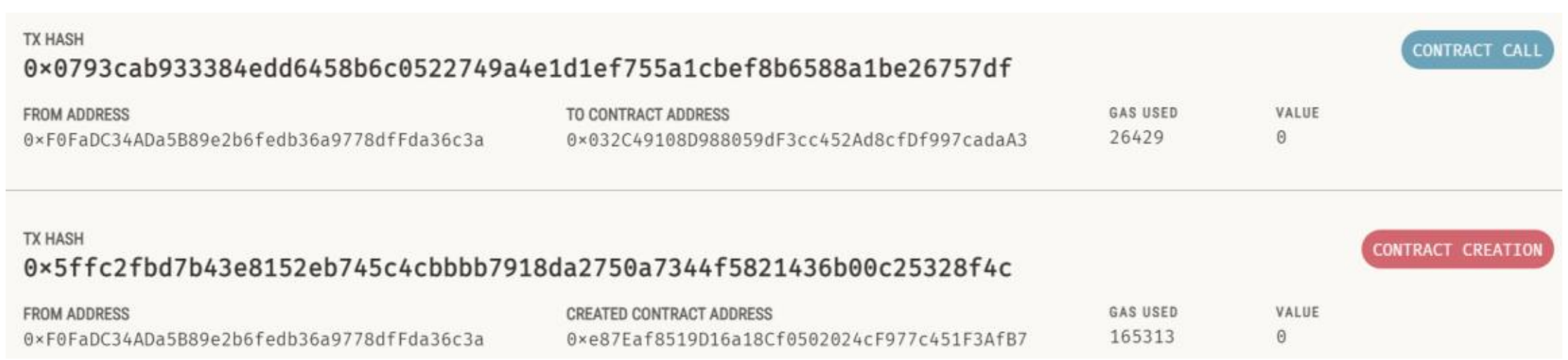

图 6、Ganache 环境交易记录

### 4.2 智能合约 fedLearn 开发

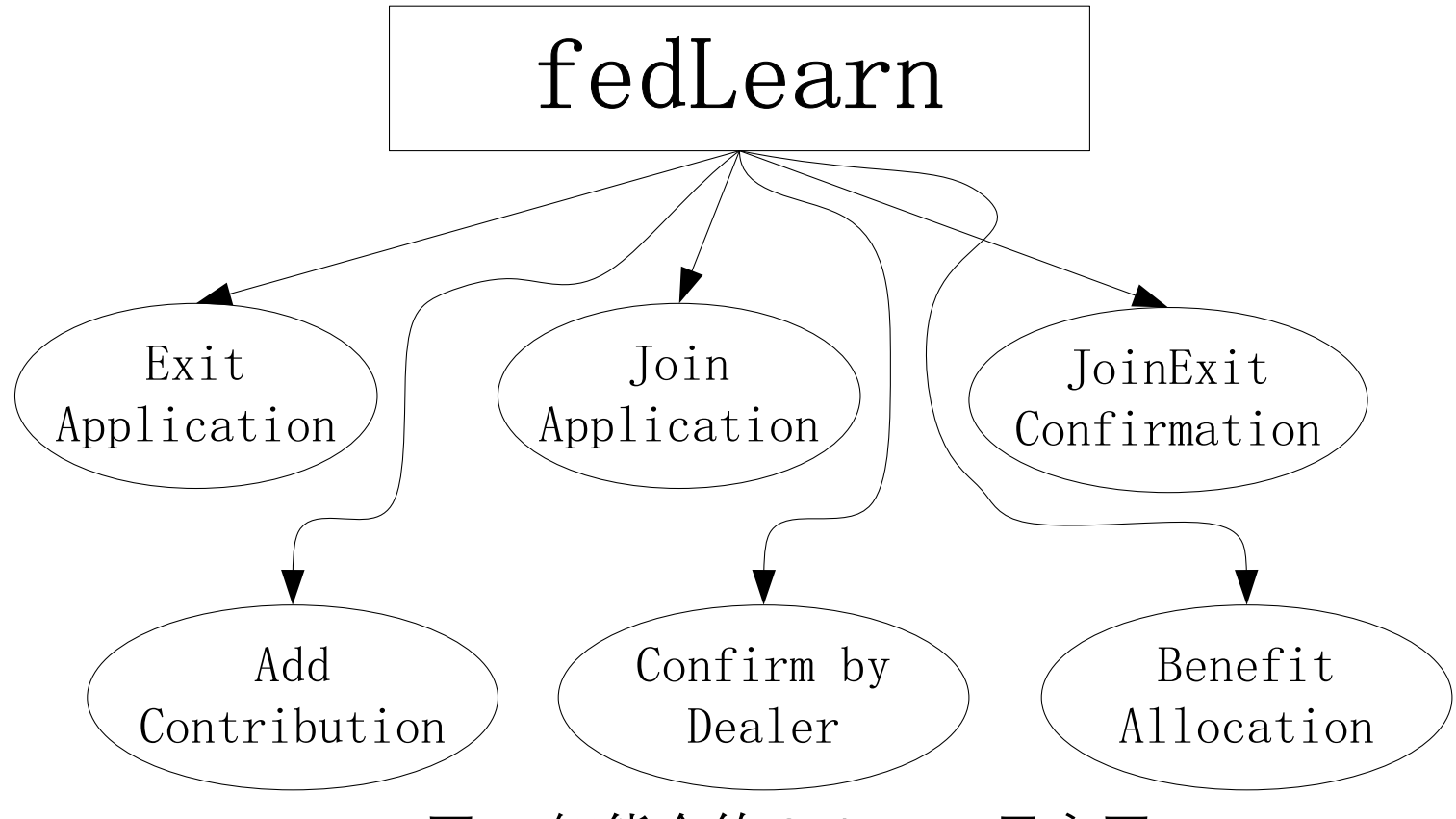

图 7. 智能合约 fedLearn 示意图

合约名为 fedLearn:

```
contract fedLearn{
        bytes32[] public participantNames;
        mapping (byte32 => uint8) public contributions;
        mapping (byte32 => bool) confirmed;
......
```

其中数组 participantNames 用来保存参与方的名字，用映射 contributions 保存参与方的贡献值，这是日后利益分配的基础，由参与方提供自己的贡献值，但不能立即生效，而是要由 Dealer 对参与方的贡献值进行检查和确认，确认的结果保存与映射 confirmed 之中，True 表示此参与方的贡献值得到 Dealer 的确认可以参与后续利益分配，False 则表明贡献未得到确认不能参与后续的利益分配。

合约中有 JoinApplication 函数，有意参与的各方可以向 Dealer 提出加入申请；而想要退出的参与方可调用合约中的 ExitApplication 函数来向 Dealer 申请退出：

```
function JoinApplication(byte32 participant, string participant_info)
function ExitApplication(byte32 participant, string exit_reason)
```

申请加入时要提供参与方名称 participant，以及参与方的相关信息 participant_info，让 Dealer 进行审核；申请退出时要提供退出原因 exit_reason；收到加入或申请时，Dealer 会通过 JoinExitConfirmation 函数进行审核，Dealer 会把通过审核的加入方添加到上述参与方的数组中，让其可以参与后续的合作；而把通过审核的退出方的名字从参与方数组中删除，不再参与后续的合作和分红等：

```
function JoinExitConfirmation( )
```

合约中提供了 addContibution 函数：

```
function addContibution(byte32 participant, uint8 contribution) {
    paticipantNames.push(participant);
    contributions[participant]=contribution;
}
```

参与方可调用此函数向区块链中写入自己的名字和贡献值；之后 Dealer 可调用合约中的 confirm 函数对参与方写入的贡献值进行确认：

```
function confirm (byte32 participant, uint8 contribution)
```

```
        require (“call by the dealer”);
        if contribution == contributions[participant]
            confirmed[participant] = true;
        else:
            confirmed[participant] = false;
    }
```

首先通过 require 语句确保只有 Dealer 能够调用此函数，而后比较 Dealer 认可的贡献值与参与方自己写入的贡献值，相同的话则更新 confirmed 映射为 True，否则为 False。

利益分配函数 benifitAllocate，必须由 Dealer 调用，先计算各个参与方的贡献比例，据此划分收益，并通过 send money 函数将收益发送给相应的参与方，注意只用经过 Dealer 确认过的贡献方和贡献值（即 confirmed 为 True）才能参与分配：

```
    function benifitAllocate(uint8 money){
        require(“called by the dealer”)
        uint8 sum=0
        for(uint8 i=0; i<paticipants.length;i++) {
            if (confirmed(participants[i]) sum+=contribution[paticiaptns[i]];
          }
        for(uint8 i=0; i<paticipants.length;i++) {
        if(confirmed(participants[i])                                        {
            uint8 percent = contributions[particiants[i]] * 100 / sum
            sendmoney(participants[i], money * percent / 100)
        }
```

**5. 结束语**

今天的大数据时代，大量的数据带有隐私性质被保存于各人或单位的本地数据库之中，由于隐私保护问题很难进行充分的挖掘和利用，而联邦学习技术的出现从一定程度上解决了这一问题，因其可以在保护隐私的前提下进行数据挖掘、机器学习等操作。

联邦学习中有多个参与方，各自数据的质量和数量不同，贡献各有不同，本文探讨如何保护参与联邦学习的各方的权益，提出一种基于去中心化的区块链和智能合约的解决方案；方案大致流程如下，Dealer 发布招募参与者的帖子，说明要什么数据、开发什么产品、最终产品的收费标准、利益如何分成，以及相关神经网络参数等，并发布智能合约的地址、调用方法与参数等；而有兴趣要参加的数据持有者可报名参与，根据帖子的要求，根据联邦学习的相关操作本地训练局部神经网络模型，然后把训练所得的局部神经网络模型参数发送给 Dealer，同时调用帖子中的智能合约接口向区块链上写入自己的贡献值，作为日后参与产品收益分成的依据。

本文方案还有一些不足之处，比如还有一个中心 Dealer，在后续研究中拟进一步去中心化，通过相关密码学协议取代 Dealer，构建无中心的联邦机器学习，完全 P2P 模式，成员可以随意加入退出，产品以比特币收益，达到某个标准就进行自动的利益分配；另一可能的研究方向是更紧密结合区块链和联邦学习的分布式操作，设计更高效、更安全、更简洁的方案。

**参考文献：**

[1]. Brendan McMahan, Moore Eider, Ramage Daniel, etal. Communication-Efficient Learning of Deep Networks from Decentralized Data[A]//Singh A, Zhu J. Proceedings of

Machine Learning Research: PMLR, 2017: 1273-1282.
[2]. 孙钰，严宇，崔剑，等. 联邦学习深度梯度反演攻防研究进展[J]. 电子与信息学报, 2024.
[3]. 魏立斐，张无忌，张蕾，等.基于本地差分隐私的异步横向联邦安全梯度聚合方案[J].电子与信息学报, 2024.
[4]. 李敏，肖迪，陈律君.兼顾通信效率与效用的自适应高斯差分隐私个性化联邦学习[J].计算机学报, 2024.
[5]. 谭作文，张连福. 机器学习隐私保护研究综述[J]. 软件学报, 2020, 31(07): 2127-2156.
[6]. 穆旭彤，程珂，宋安霄，等. 抗拜占庭攻击的隐私保护联邦学习[J]. 计算机学报, 2024.
[7]. 张仁斌，崔宇航，张子石.基于 β-VAE 的联邦学习异常更新检测算法[J].计算机应用研究, 2024.
[8]. 张帅华，张淑芬，周明川，等.基于半监督联邦学习的恶意流量检测模型[J].计算机应用, 2024.
[9]. 陆枫，李炜，顾琳，等.基于迭代协作学习框架的信誉医学参与方选择[J].计算机研究与发展, 2024.
[10]. 韩沛秀，孙卓，刘忠波，等.基于个性化联邦学习的异构船舶航行油耗预测[J].计算机集成制造系统, 2024.
[11]. 张新，谭学泽，王小艺，等.基于区块链和联邦学习的食品全程全息风险信息可信共享模式[J]. 食品科学, 2024.
[12]. 崔腾，张海军，代伟. 基于分布共识的联邦增量迁移学习[J]. 计算机学报, 2024.
[13]. 董成荣，姚俊萍，李晓军，等.面向分布式复杂数据样本的联邦语义分割方法研究综述[J]. 计算机应用研究, 2024
[14]. 琚文胜,张世红,桑戟南,等.互联网医疗个人健康医疗数据确权授权分析研究[J].中国卫生信息管理杂志,2023,20(06):904-911.
[15]. 王泽.以“数据二十条”内在机理为导向，论数据要素安全与发展实践路径[J].网络安全技术与应用,2024,(01):66-70.
[16]. 李国昊,梁永滔,苏佳璐.数据确权背景下电商平台大数据“杀熟”治理分析[J].运筹与管理,2023,32(11):227-232.
[17]. 汪晓梅,黄科满,邵瑞江.支撑数据要素流通的数据交易平台的关键技术设计与实践[J].新型工业化,2023,13(11):40-50.
[18]. 马费成,熊思玥,孙玉姣,等.数据分类分级确权对数据要素价值实现的影响[J].信息资源管理学报,2024,14(01):4-12.DOI:10.13365/j.jirm.2024.01.004.
[19]. 齐爱民,杨煜.人工智能生成内容确权模式的嬗变与数据财产权的确立[J].苏州大学学报(哲学社会科学版),2024,45(01):88-99.DOI:10.19563/j.cnki.sdzs.2024.01.009.
[20]. 袁勇,王飞跃.区块链技术发展现状与展望[J].自动化学报,2016,42(04):481-494.DOI:10.16383/j.aas.2016.c160158.
[21]. 邵奇峰,金澈清,张召,等.区块链技术:架构及进展[J].计算机学报,2018,41(05):969-988.
[22]. 许鑫,叶丁菱.数智时代情报学与情报工作的发展透视[J].情报学报,2022,41(10):1100-1110.
[23]. 郑荣,高志豪,魏明珠,等.基于联盟区块链的政府数据协同治理平台框架研究——以全国碳排放权交易市场为例[J].情报学报,2022,41(10):1071-1084.
[24]. Y. Miao, Z. Liu, H. Li, K. -K. R. Choo and R. H. Deng, "Privacy-Preserving Byzantine-Robust Federated Learning via Blockchain Systems," in IEEE Transactions on Information Forensics and Security, vol. 17, pp. 2848-2861, 2022, doi: 10.1109/TIFS.2022.3196274.

[25]. 米波,翁渊,黄大荣,等.基于区块链共识激励机制的新型联邦学习系统[J].信息安全学报,2024,9(01):15-32.DOI:10.19363/J.cnki.cn10-1380/tn.2024.01.02.

[26]. 张新,谭学泽,王小艺,等.基于区块链和联邦学习的食品全程全息风险信息可信共享模式[J].食品科学,2024,45(15):1-12.

[27]. V. S. Chundawat, A. K. Tarun, M. Mandal and M. Kankanhalli, "Zero-Shot Machine Unlearning," in *IEEE Transactions on Information Forensics and Security*, vol. 18, pp. 2345-2354, 2023.